\documentclass[12pt]{article}
\pdfoutput=1
\usepackage{mathtools,amssymb,mathrsfs,microtype,tikz,ytableau}
\usepackage{tikz-cd}
\usepackage{tabularx,ragged2e}
\usepackage[utf8]{inputenc}
\usepackage{float}
\usepackage{xcolor}
\usepackage{empheq}
\usepackage{amsmath}
\usepackage{slashed}
\usepackage{caption}
\usepackage{dsfont}
\usepackage{verbatim}
\usepackage{tcolorbox}
\usepackage{subfig}
\usepackage{amsfonts}
\usepackage{graphicx}
\usepackage{physics}
\usepackage{slashed}
\usepackage[colorlinks=true]{hyperref}
\def\Circlearrowleft{\ensuremath{%
  \rotatebox[origin=c]{180}{$\circlearrowleft$}}}

\newif\ifnotoc\notocfalse
\newif\ifemailadd\emailaddfalse
\newif\iftoccontinuous\toccontinuousfalse
\newif\ifnatbibsort\natbibsorttrue

\relax

\ifnatbibsort\RequirePackage[numbers,sort&compress]{natbib}\else\RequirePackage[numbers,compress]{natbib}\fi

\ytableausetup{centertableaux,boxsize=.5em}
\makeatletter\@addtoreset{equation}{section}\makeatother

\renewcommand{\title}[1]{\vbox{\center\LARGE{#1}}\vspace{5mm}}
\renewcommand{\author}[1]{\vbox{\center\large#1}\vspace{5mm}}

\newcommand{\beq}{\begin{equation}\begin{aligned}}
\newcommand{\eeq}{\end{aligned}\end{equation}}

\newcommand{\del}{\delta}

\newcommand{\Sig}{\Sigma}
\newcommand{\ov}{\over}
\newcommand{\lr}{\left \langle}
\newcommand{\rr}{\right \rangle}
\newcommand{\sgn}{\text{sgn}}
\newcommand{\pa}{\partial}
\newcommand{\mcal}{\mathcal}

\begin{document}
 
\begin{titlepage}
\begin{center}
\hfill {\tt CERN-TH-2019-206}\\
\vspace{1mm}

\title{
 {\LARGE Exact four point function for large $q$ SYK from Regge theory}}
\vspace{4mm}

Changha Choi,${}^{a,b}$\footnote{\href{mailto: changhachoi@gmail.com}
{\tt changhachoi@gmail.com}}
M\'ark Mezei${}^{b}$,\footnote{\href{mailto: mmezei@scgp.stonybrook.edu}
{\tt mmezei@scgp.stonybrook.edu}}
G\'abor S\'arosi${}^{c}$\footnote{\href{mailto: saro.gabor@gmail.com}
{\tt saro.gabor@gmail.com}}

\vskip 8mm

{\it
${}^{a}$Physics and Astronomy Department,\\
Stony Brook University, Stony Brook, NY 11794, USA\\
\vspace{0.2cm}
${}^{b}$Simons Center for Geometry and Physics,\\
Stony Brook University,
Stony Brook, NY 11794, USA\\
\vspace{0.2cm}
${}^{c}$CERN, Theoretical Physics Department, 1211 Geneva 23, Switzerland
\vspace{0.2cm}
 }

\end{center}

\vspace{5mm}
\abstract{
\noindent Motivated by the goal of understanding quantum systems away from maximal chaos, in this note we derive a simple closed form expression for the fermion four point function of the large $q$ SYK model valid at arbitrary temperatures and to leading order in $1/N$. The result captures both the large temperature, weakly coupled regime, and the low temperature, nearly conformal, maximally chaotic regime of the model. The derivation proceeds by the Sommerfeld-Watson resummation of an infinite series that recasts the four point function as a sum of three Regge poles. The location of these poles determines the Lyapunov exponent that interpolates between zero and the maximal value as the temperature is decreased. Our results are in complete agreement with the ones by Streicher \cite{Streicher:2019wek} obtained using a different method.
}
\vfill

\end{titlepage}

{\hypersetup{linkcolor=black}
\tableofcontents
\thispagestyle{empty}
}

\section{Introduction}
\label{sec:intro}
\setcounter{footnote}{0} 

Recent years have seen great progress in the understanding of many-body quantum chaos. These developments were greatly aided by explicit results in holographic gauge theories, where the gravitational description helped uncover the relation between the butterfly effect, operator growth, and out of time order correlation functions \cite{Shenker:2013pqa,Roberts:2014isa,Shenker:2014cwa}. The introduction of the Sashdev-Ye-Kitaev (SYK) model \cite{kitaev2014hidden,Polchinski:2016xgd,Maldacena:2016hyu} then provided the first solvable model of quantum chaotic dynamics. Its low energy description in terms of the Schwarzian effective theory led to the understanding of the NCFT$_1$ universality class of quantum dynamics, and its holographic dual JT gravity description \cite{Jensen:2016pah,Maldacena:2016upp}.

Most of the focus on the literature has been on maximally chaotic theories: semiclassical gravity, the near conformal low temperature limit of SYK, and two-dimensional CFTs with very sparse spectra.\footnote{The conditions for the identity Virasoro block dominance for the out of time order four point functions are not yet fully understood, see however \cite{Mezei:2019dfv} in the Lyapunov regime.} It is interesting to also explore chaotic theories that do not exhibit maximal chaos. By understanding their properties we may appropriately generalize the new theoretical structures uncovered in the study of maximally chaotic systems. Some promising directions are:
\begin{itemize}
\item Understanding the structure of out of time order correlation functions inside the butterfly cone. In the maximally chaotic case the spatial structure is found to be very simple and it would be interesting to understand it in the generic case
\beq
\text{OTOC}(t,x)=1-{\#\over N^2}\, \exp\left[{2\pi T} \left(t-{\abs{x}/v_B}\right)\right]\quad  \rightsquigarrow \quad 1-{\#\over N^2}\, \exp\left[\lambda\left(\abs{x}\over t\right)\,t\right]\,,
\eeq 
where we introduced the velocity dependent Lyapunov exponent $\lambda\left(v\right)$ \cite{Xu:2018xfz,Khemani:2018sdn,Mezei:2019dfv}.\footnote{The more familiar Lyapunov exponent is obtained by setting $v=0$: $\lambda_L=\lambda\left(0\right)$.}
\item Several tantalizing coincidences have been noticed in maximally chaotic theories: the relation between energy diffusion and chaos \cite{Blake:2016sud} and the pole skipping phenomenon \cite{Blake:2017ris}. It would be interesting to explore, in what sense (if at all) these remain true away from maximal chaos \cite{Grozdanov:2018kkt}.
\item The generalization of the Schwarzian, a geometric action capturing the reparametrizations of the time variable of the theory, away from maximal chaos to obtain the so-called scramblon action is an outstanding challenge \cite{Kitaev:2017awl}.\footnote{The exchange of the scramblon (or pomeron) would generate non-maximal chaotic growth. It is unclear whether the same scramblon mode would also be responsible for energy transport, or there is a separate  hydrodynamic mode, and the two become equal only at maximal chaos.  } There have been some effective actions proposed in the literature that capture some or all of the aforementioned phenomena \cite{Blake:2018leo,Cotler:2018zff,Haehl:2018izb,Haehl:2019eae}, but currently it is not known how to generalize them away from maximal chaos. 

\end{itemize}

To make progress in these directions, it seems helpful to have an explicit example, where some of these ideas can be tested. 
In this note, we present such an example: the fermion four point function of the large $q$ SYK model valid at arbitrary temperatures given in \eqref{eq:toc} and \eqref{eq:otoc} in terms of elementary functions. It was noted early on that in the large $q$ limit the SYK model simplifies considerably, and one can obtain analytic expressions away from the near conformal limit \cite{Maldacena:2016hyu}. These were given in the literature in the form of an infinite sum. In the process of exploring the aforementioned directions, especially the possibility to write a scramblon effective action, we have managed to resum the infinite series using the Sommerfeld-Watson resummation technique familiar from Regge theory \cite{watson1918diffraction,Cornalba:2007fs,Costa:2012cb}. The answer is given in terms of very simple contributions of three Regge poles: in the time ordered limit, only the one at the origin contributes, and hence there is no exponential growth, while for the out of time order case all three of them contribute. These Regge poles then determine the exponential growth of the OTO four-point function to be $\lambda_L=2\pi T\, v$, where $v\in(0,1)$ is the effective coupling constant, a result already obtained in \cite{Maldacena:2016hyu} using the retarded kernel approach.

It would be very interesting to make progress on any of the aforementioned important open problems using these explicit examples. A generalization to the large $q$ SYK chain \cite{Gu:2016oyy} would be desirable, since it could provide spatial locality missing from the single SYK dot. Very recently, in \cite{Streicher:2019wek} Streicher has obtained the same four point function that we derive here using a seemingly very different, elegant method. Our results are in complete agreement. We decided to write this note to provide an alternative, Regge theory perspective on the four point function.

\section{Review of the large $q$ SYK model} \label{sec:review}

In this review section, we largely follow \cite{Maldacena:2016hyu}. The SYK model model is a quantum mechanical system with $N$ Majorana fermions satisfying $\{ \psi_i,\psi_j\}=\del_{ij}$ coupled through an interacting Hamiltonian with disorder. The Hamiltonian is characterized by a positive even integer $q$, which represents the number of fermions that randomly interact with each other at tree-level as follows:

\beq \label{eq:model}
H=(i)^{{q\ov 2}} \sum_{1\leq i_1 <\dots <i_q\leq} j_{i_1 \dots i_q}\psi_{i_1}\dots \psi_{i_q},
\eeq
where $j$ is from a random Gaussian distribution normalized as $\langle j_{i_1 \dots i_q}^2\rangle={J^2 (q-1)!\ov N^{q-1}}={2^{q-1} \mathcal J^2 (q-1)!\ov q N^{q-1} }$. As we will see below, the large $q$ limit of the SYK model is well-defined when we keep the parameter $\mathcal J$ fixed.

The main basic feature of the SYK model is that the Schwinger-Dyson equation of the Euclidean time-ordered fermion propagator $\lr T\{\psi_i(\tau)\psi_j(0)\}\rr\equiv G(\tau)\del_{ij} $ has a simple leading large $N$ expression because of the dominance of the melonic Feynman-diagrams:

\beq \label{eq:Schwinger-Dyson}
~&{1\ov G(\omega)}=-i\omega-\Sigma(\omega)
\\& \Sigma(\tau)=J^2 G(\tau)^{q-1}.
\eeq

Now the large $q$ limit of the SYK model is defined as we take $q\rightarrow \infty$ and obtain physical quantities in an ${1\ov q}$ expansion. It is important that the large $q$ limit is taken after we take the large $N$ limit above, which allows one to keep having an simplicity of the melonic dominance, see \cite{Berkooz:2018qkz} when corrections of the type $q/N$ are kept. Next, we discuss about the general characteristics of the leading large $N$ two and four point function of the SYK model.

\subsection{Two point function}

In the large $q$ limit, we fix $\mathcal J \equiv 2^{{1-q\ov 2}}\sqrt{q}J$ as a constant, so that the full propagator has a well-defined ${1\ov q}$ expansion with respect to the free propagator $G_{\text{free}}(\tau)={1\ov 2}\text{sgn}(\tau)$ as follows:
\beq \label{eq:propagator}
~&G(\tau)={1\ov 2}\text{sgn}(\tau)\left[ 1+{1\ov q} g(\tau)+O\left({1\ov q^2}\right)\right]
\\& \Sig(\tau)={\mathcal J^2 \ov q}\text{sgn}(\tau) e^{g(\tau)}\left[1+O\left({1\ov q}\right)\right]
\eeq

The highlight of the large $q$ SYK two point function is that the first non-trivial correction to the propagator $g(\tau)$ is exactly solvable, once we plug the expansion \eqref{eq:propagator} into the Schwinger-Dyson equation \eqref{eq:Schwinger-Dyson}. This becomes a linear ODE 
\beq
\partial_\tau^2 [\text{sgn}(\tau) g(\tau)]=2\mathcal J^2 \text{sgn}(\tau) e^{g(\tau)},~ g(0)=g(\beta)=0,
\eeq
which has the following unique solution with appropriate thermal boundary condition for inverse temperature $\beta$
\beq
~&e^{g(\tau)}=\left[{\cos({\pi v \ov 2} ) \ov \cos (\pi v ({1\ov 2}-{\vert t \vert \ov \beta}))} \right]^2,
\eeq
where $0\leq v\leq 1$ is implicitly defined by $\beta \mathcal J={\pi v  \ov cos{\pi v \ov 2}}$. $v$ is a monotonic function of $\beta J$ where the low-temperature (or strong-coupling) limit becomes $v\rightarrow 1$, while the high-temperature (or weak-coupling) limit becomes $v\rightarrow 0$.

\subsection{Ladder diagram and the four point function.}

The next basic physical quantity that contains dynamical information about the full theory is the four point function of the Majorana fermions. Let's consider the ${1\ov N}$ expansion of the disorder averaged four point function

\beq \label{eq:4pt def}
{1\ov N}\sum_{i,j=1}^N \lr T\{ \psi_i(\tau_1)\psi_i (\tau_2) \psi_j(\tau_3)\psi_j(\tau_4) \}\rr=G(\tau_{12})G(\tau_{34})+{1\ov N}\mathcal F(\tau_1 ,\tau_2,\tau_3,\tau_4)+O\left({1\ov N^2}\right)
\eeq

In the large $N$ limit, $\mathcal F$ contains the leading non-trivial information. As in the case of the two point function, one can obtain a simple Schwinger-Dyson equation for $\mathcal F$  from the resummation of ladder diagrams ($\tau_{ij}\equiv \tau_i-\tau_j$) \cite{Polchinski:2016xgd,Maldacena:2016hyu}
\beq \label{eq:ladder 4pt}
~&\mathcal F= {1\ov 1-K} \mathcal F_\text{dis}
\\& \mathcal F_\text{dis}= -G(\tau_{13}) G(\tau_{24}) +G(\tau_{14})G(\tau_{23})
\\&K=-J^2 (q-1) G(\tau_{13}) G(\tau_{24}) G(\tau_{34})^{q-2}
\eeq

The interpretation of eq. \eqref{eq:ladder 4pt} is following. $\mathcal F_\text{dis} $ is the disconnected four point function originating from the contraction between $i$ and $j$ fermions in \eqref{eq:4pt def}. Now the large $N$ diagrammatic reveals that the only contribution to the $O({1\ov N})$ four point function arises from the so-called ladder diagrams with the $n$ ($n=0,1,2,...$) parallel rungs each made from  $q-2$ internal contractions between the two interaction vertices. Addition of an extra rung can be interpreted as an insertion of the kernel $K$ defined as 
\beq
K(\tau_1, \tau_2 ; \tau_3,\tau_4)=-J^2(q-1) G(\tau_{13})G(\tau_{24}) G(\tau_{34})^{q-2},
\eeq
and the resummation of the whole ladder diagram can be represented as a geometrical series which becomes the first line of \eqref{eq:ladder 4pt}. A notable feature of the large $q$ limit is the simplicity of the kernel $K$ as follows \cite{Maldacena:2016hyu}

\beq \label{eq:kernel}
~&K(\tau_1, \tau_2;\tau_3,\tau_4)=-{v^2 \ov 2 \cos^2 \left(\pi v \left({1\ov 2}-{\vert \tau_{34} \vert \ov \beta}\right)\right)}\sgn(\tau_{13})\sgn(\tau_{24})
\\& K^{-1}(\tau_1, \tau_2;\tau_3,\tau_4)=-{2 \cos^2 \left(\pi v \left({1\ov 2}-{\vert \tau_{34} \vert \ov \beta}\right)\right) \ov v^2 }\pa_{\tau_1}\pa_{\tau_2}
\eeq
Alternatively, one may understand \eqref{eq:ladder 4pt} as the two point function of fluctuations in the bilocal effective action of the model. This action becomes a certain Lorentzian Liouville theory in the large $q$ limit \cite{Cotler:2016fpe}, whose fluctuations are also governed by \eqref{eq:kernel}.

From now on we set the temperature to be $\beta=2\pi$. Before obtaining the final differential equation that determines $\mathcal F$, let's discuss  the domain and the symmetry of the four point function. Because of the Euclidean thermal circle, we can restricts the dynamical time variable to be within the interval $\tau\in [0,2\pi)$. Then Fermi statistics along the thermal circle allows us to extend the four point function to any real value of $\tau\in\mathcal R$. At this stage it is convenient introduce the symmetric coordinates $(x,y;x',y')\equiv (\tau_1-\tau_2, {\tau_1+\tau_2\ov 2}; \tau_3-\tau_4,{\tau_3+\tau_4\ov 2})$ which will become clear as we proceed. Then we can define the fundamental domain of the four point function as $D:0\leq x,x'<2\pi, ~0\leq y,y'\leq 2\pi$ because of the following symmetry constraints

\beq \label{eq:sym}
~&\mathcal F(x,y;x',y')=-\mathcal F(-x,y;x',y'),~ \mathcal F(x,y;x',y')=-\mathcal F(x,y;-x',y'),
\\&\mathcal F(x,y;x',y')=\mathcal F(2\pi-x,y\pm \pi;x',y')=\mathcal F(x,y;2\pi -x',y'\pm \pi) ,
\eeq

Now we are ready to specify the four point function as a problem of finding the Green's function for a PDE. First we recast the Schwinger-Dyson equation \eqref{eq:ladder 4pt} as $(K^{-1}-1)\mathcal F=K^{-1}\mathcal F_\text{dis}$. Together with \eqref{eq:kernel}, we get the PDE for the resummed four point function $\mathcal F$ subject to the symmetry constraints \eqref{eq:sym} in the domain $D$ (here we impose the further restriction $0\leq y'\leq \pi$ using \eqref{eq:sym}):

\beq \label{eq:green}
\left( -{\pa_y^2\ov 4}+v^2\pa_{\tilde x}^2-{v^2 \ov 2 \sin^2({\tilde x\ov 2}) }\right)\mathcal F(x,y;x',y')=\delta(y-y')\delta(x-x')+\delta(y-y'-\pi)\delta(2\pi-x-x'),
\eeq
where we changed to the variable $\tilde x\equiv vx+(1-v)\pi$, with $(1-v)\pi<\tilde x<(1+v)\pi$. In the next section, we solve the above Green's equation \eqref{eq:green} in a closed form, using only elementary functions.

\section{Exact leading large $N$ four point function} \label{sec:exact}

The first step of solving \eqref{eq:green} is to identify the eigenfunctions of the differential operator $\mathcal L\equiv \left( -{\pa_y^2\ov 4}+v^2\pa_{\tilde x}^2-{v^2 \ov 2 \sin^2({\tilde x\ov 2}) }\right)$. Separation of  variables together with the symmetry constraints \eqref{eq:sym} gives the following eigenfunctions $\Psi_{n,m}(x,y)$ (we use the original variable $x$ for the eigenfunctions even though we used $\tilde x$ in the differential equation.):

\beq
~& \mathcal L \Psi_{n,m}(x,y)={n^2-m^2\ov 4}\Psi_{n,m}(x,y)
\\&\Psi_{n,m}(x,y)=\begin{cases} e^{i n y} \psi^e_m(x) & n\in 2\mathbb Z,  ~m\in \mcal M^{e}\equiv \{m\geq0\vert \psi_m^e(0)=0\}  \\ e^{iny} \psi^o_m(x) &n\in 2\mathbb Z+1, ~m\in \mcal M^{o}\equiv \{m\geq0 \vert \psi_m^o(0)=0\} \end{cases}
\eeq
where $\psi_m^{e}(x)$ and $\psi_m^{o}(x)$ are the eigenfunctions of $\left[v^2\pa_{\tilde x}^2-{v^2 \ov 2 \sin^2({\tilde x\ov 2}) }\right]$ with eigenvalue ${m^2\ov 4}$ that under $x\rightarrow 2\pi-x$ are even and odd respectively. The explicit form of the eigenfunctions is the following (see e.g. \cite{lekner2007reflectionless})
\beq
~&\psi^e_m(x)=\frac{m \cos \left(\frac{1}{2} m (\pi -x)\right)}{v}+\sin \left(\frac{1}{2} m (\pi
   -x)\right) \tan \left(\frac{1}{2} v (\pi -x)\right)
\\&\psi^o_m(x)=\frac{m \sin \left(\frac{1}{2} m (\pi -x)\right)}{v}-\cos \left(\frac{1}{2} m (\pi
   -x)\right) \tan \left(\frac{1}{2} v (\pi -x)\right)
\eeq
The same eigenfunctions were given as hypergeometric functions in \cite{Maldacena:2016hyu}.

The variable $m$ is not aribtary but determined by the Dirichlet boundary condition $\psi_m^{e,o}(0)=0$ which follows from the first line of the symmetry constraints \eqref{eq:sym}. We restrict it to be $m\geq 0$ to avoid a double counting since $\psi_{-m}^e(x)=-\psi_{m}^e(x),~\psi_{-m}^o(x)=\psi_{m}^o(x)$. The index sets ${\cal M}^{e,o}$ are  explicitly given by:

\beq
~& \mcal M^e:~\psi_m^e(0)=0~\Leftrightarrow~ m \cot \left(\frac{m\pi }{2} \right)+v \tan \left(\frac{v\pi }{2} \right)=0
\\& \mcal M^o:~ \psi_m^o(0)=0~\Leftrightarrow~m \tan \left(\frac{m\pi }{2} \right)-v \tan \left(\frac{v\pi }{2} \right)=0
\eeq

We note that because of the triviality of $\psi_0^e(x)$ and $\psi_v^0(0)$, the allowed modes $m\in \mcal M^e\cup \mcal M^o$ are bounded from below as $\vert m\vert > v$. For example, the first few modes in the case of $v={3\ov 5}$ correspond to $\mcal M^e=\{1.35, 3.16, 5.10,7.07,9.06,\dots\},~\mcal M^o=\{2.23,4.13,6.09,8.06,10.05,\dots\}$ where the first five points of $\mcal M^e\cup \mcal M^o$ are shown in the figure \ref{fig:Contour}. 

Let's define the normalization $\mathcal N^{e,o}\equiv \int_{x=0}^{2\pi} \psi^{e,o}_m(x)^2 dx$, which explicitly reads as
\beq
\mathcal N_m^e={(m-v)(m+v)(m\pi-\sin(m\pi) ) \ov mv^2 }, \quad \quad \mathcal N_m^o={(m-v)(m+v)(m\pi+\sin(m\pi) ) \ov mv^2 }.
\eeq
We can then represent $\mathcal F$ as an infinite summation over the eigenfunctions:

\beq \label{eq:sum}
\mathcal F(x,y;x',y')=& \sum_{n\in 2\mathbb Z , m\in \mcal M^e}\frac{1}{2\pi \mathcal N_m^e}\frac{8e^{-in (y-y')} \psi^e_m(x)  \psi^e_m(x')}{n^2-m^2}
\\&+\sum_{n\in 2\mathbb Z +1, m\in \mcal M^o} \frac{1}{2\pi\mathcal N_m^o}\frac{8e^{-in (y-y')} \psi^o_m(x)  \psi^o_m(x')}{n^2-m^2}
\eeq
Summing over the $n$ can be done easily \cite{Maldacena:2016hyu}, and leads to the following expression 

\beq \label{eq:sumeo}
~&\mathcal F(x,y,x',y')=\mathcal F^e(x,y,x',y')+\mathcal F^o(x,y,x',y')
\\&\mathcal F^e(x,y,x,y')=\sum_{ m\in \mcal M^e}-\frac{2}{m \sin(m\pi )}(\cos (m (y-y'-\pi))+\cos(m(\vert y-y'-\pi \vert -\pi)))\frac{ \psi^e_m(x)  \psi^e_m(x')}{\mathcal N^e(m)}
\\&\mathcal F^o(x,y,x',y')=\sum_{ m\in \mcal M^o}-\frac{2}{m \sin(m\pi )}(\cos (m (y-y'-\pi))-\cos(m(\vert y-y'-\pi \vert -\pi))) \frac{ \psi^o_m(x)  \psi^o_m(x')}{\mathcal N^o(m)}
\eeq
where we have separated the contribution of even and odd center of mass energy (recall that $y={\tau_1+\tau_2\ov 2}$ can be considered as a center of mass coordinate of the two fermions at $\tau_1$ and $\tau_2$).
 
  \subsection{General remarks on the Sommerfeld-Watson transformation}\label{sec:sommerfeld-watson}
 
Here we explain the essential technical steps in the resummation of the series \eqref{eq:sumeo} the using Sommerfeld-Watson transformation. The Sommerfeld-Watson transformation is an ubiquitous complex analysis technique\footnote{The technique is usually attributed to Watson \cite{watson1918diffraction}.} in many areas of physics ranging from electromagnetism to conformal field theory. The essential idea is to write an infinite sum with analytic summands in terms of a simple contour integral involving finite number of poles which are sometimes called `Regge Poles'. This is because in the Regge theory of scattering amplitudes, one could associate the Regge poles with bound states or resonances.

A pedagogical example of the Sommerfeld-Watson transformation can be found in standard thermal field theory, where in the imaginary time formulation one often encounters the infinite summation over the evenly spaced Matsubara frequencies $\omega_n=2\pi n$(we set the temperature to be 1). For example, let's consider the sum $\sum_{n\in \mathbb Z} f(\omega_n)$ and suppose $f(\omega)$ has a nice asymptotic properties in the upper and the lower half plane of the complexified $\omega$. Then Cauchy's theorem allows one to recast the infinite series in terms of single contour integral with the contour circling the integers as follows: 
\beq \label{eq:residue}
\sum_{n\in \mathbb Z} f(\omega_n)={1\ov 2\pi }\oint_{C}{f(\omega)s_+(\omega )d\omega},
\eeq
 where $s_+(\omega)={1\ov e^{i \omega}-1}$. When $f(\omega)$ has finite number of the poles $\zeta_k$, we can obtain the closed form expression as 
 \beq \sum_{n\in \mathbb Z} f(\omega_n)=-{i }\sum_{k} \text{Res}({f(\omega)s_+(\omega)},\zeta_k).
 \eeq
  Actually, this deformation makes sense whenever ${f(\omega )s_+(\omega)}$ decays faster than $O({1\ov \omega})$. 
We note that one can equally use the ${f(\omega)s_-(\omega)}$ with $s_-(\omega)={1\ov e^{-i \omega}-1}$ provided that they converge faster than $O({1\ov \omega})$ at the complex infinity.  In many cases, the summand $f(\omega)$ doesn't satisfy the required decaying property with any single choice among $s_\pm (\omega)$, but the resummation is still possible if we divide $f(\omega)=f_+(\omega)+f_-(\omega)$ and use contour integrals with $f_+(\omega)s_+(\omega)$ and $f_-(\omega)s_-(\omega)$ respectively. We will find below that this is also the main characteristics of our summand of the four point function \eqref{eq:sumeo}.

When we try to apply this technique to our infinite series representation of the leading large $N$ four point function \eqref{eq:sumeo}, we run into several difficulties. First, the parameter $m$ that we sum over is defined to run over $m\in \mathcal M^e \cup \mathcal M^o$ which are not uniformly spaced and the set itself does not have a closed form expression. Even if we had an analytic expression, the more serious problem is finding an analog of $s_\pm(\omega)$ of the thermal field theory with similar nice analytic properties at infinity. Finally, even if we could find such a nice function and rewrite the sum as a contour integral, how could we manage to escape from an additional infinite set of simple poles coming from the ${1\ov \sin{m\pi}}$ factor in \eqref{eq:sumeo}?

Now here is the art of the present work entering in. First nice observation is that the Dirichlet boundary condition $\psi_m^{e,o}(0)=0$ allows us to replace ${1\ov \sin{m\pi}}$ with a different analytic function which has pole only at $m=0$. More surprisingly, if we also massage the $y$ dependent part of the even and odd summand using this Dirichlet boundary condition, we can unify the even and odd prefactors in the summand \eqref{eq:sumeo} in the following analytic expression

\beq
~&-\frac{2}{m \sin(m\pi )}(\cos (m (y-y'-\pi))\pm \cos(m(\vert y-y'-\pi \vert -\pi)))
\\&=\frac{ \left(v \tan \left(\frac{\pi  v}{2}\right)+i   m\right)}{m^2} e^{i m(y-y')}+\frac{ \left(v \tan \left(\frac{\pi  v}{2}\right)-i   m\right)}{m^2} e^{-im(y-y')}.
\eeq
The next non-trivial steps comes from gaining control over the parameter $m\in \mathcal M^{e}\cup \mathcal M^{o}$ by finding an analog of $s_\pm(\omega)$ corresponding to our case. It turns out that we can unify the even and odd eigenfunctions $\psi_m^{e,o}(x)$ and the corresponding normalizations $\mathcal N_m^{e,o}$ into a single expression respectively
\beq
~&\psi_m(x)=\psi^e_m(x)\psi^o_m(0)+\psi^o_m(x)\psi^e_m(0)
\\&\mathcal N(m)=\frac{(m-v) (m+v) \left(\pi  m^2+v \tan \left(\frac{\pi  v}{2}\right) \left(\pi  v \tan   \left(\frac{\pi  v}{2}\right)+2\right)\right)}{v^4}.
\eeq
Simultaneously, the zeros of the $\mcal M^e, \mcal  M^o$ are unified into $\mcal M \equiv \{ m\geq 0 \vert \psi_m(0)=0\}=\mcal M^e\cup \mcal M^o$.
This remarkable unification allows us to express the four point function \eqref{eq:sumeo} in the following way

\beq \label{eq:sumf}
\mathcal F(x,y;x',y')=\sum_{m\geq0}\left(\frac{ v \tan \left(\frac{\pi  v}{2}\right)+i   m}{m^2} e^{i m(y-y')}+\frac{ v \tan \left(\frac{\pi  v}{2}\right)-i   m}{m^2} e^{-im(y-y')}\right) \frac{ \psi_m(x)  \psi_m(x')}{\mathcal N(m)}.
\eeq

\begin{center}
\begin{figure}[!h]
\centering
\includegraphics[scale=0.75]{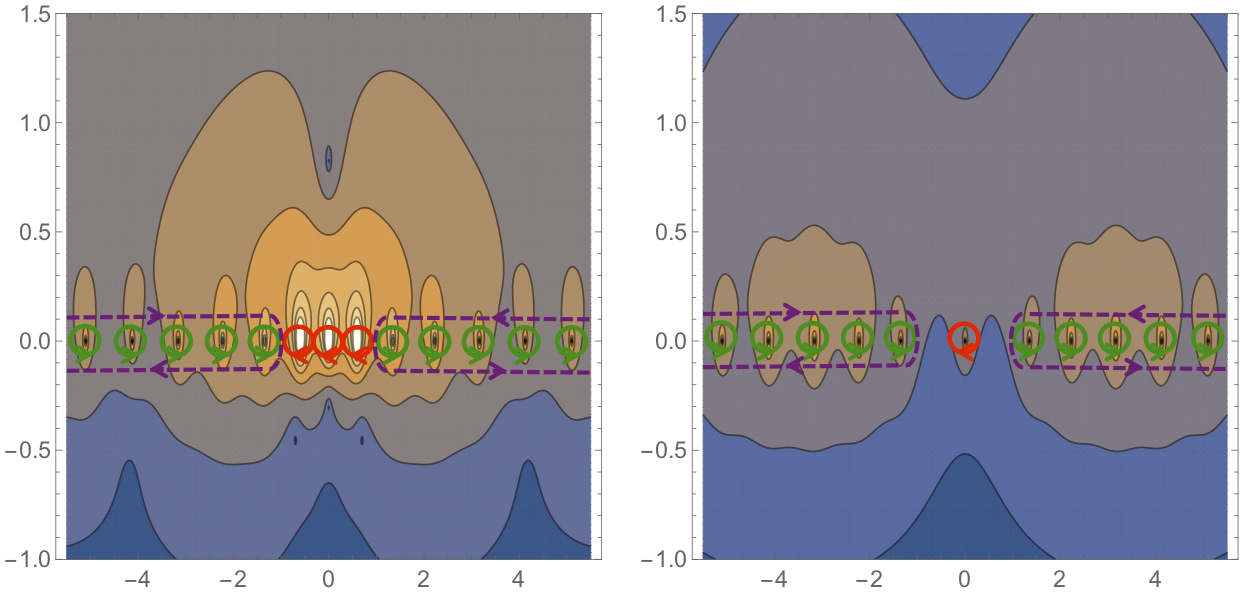}
\caption{Illustration of the contour manipulation for the Sommerfeld-Watson transformation. We first construct a meromorphic function that decays exponentially at infinity and that has poles at the values of $m$ that we want to sum over. The contribution of these poles are picked up by the green contour, which we deform into the purple dashed contour, and finally into the red contour picking up the contribution of the Regge poles. On these contour plots we plot $\abs{f_+(m)S_+(m)}^{1/4}$ for $v=3/5$, on the left for the OTOC configuration $x=x'=\pi/2,\, y=\pi/4$, while on the right for the TOC configuration $x=x'=\pi/4,\, y=\pi/2$. These figures demonstrate that for the TOC configuration the $m=\pm v$ Regge poles disappear, that the arc contributions at infinity can be dropped and that there are no additional poles or cuts on the complex $m$ plane that our analysis would have missed. \label{fig:Contour}}
\end{figure}
\end{center}

Finally, the along of $s_\pm(m)$ can be obtained starting from the observation that ${\pa_m \psi_m(x) \ov \psi_m(x)}$ has a simple pole with unit residue at $m\in \mcal M$ which allow us to transform series into a contour integral. But it turns out that the asymptotic properties are not good enough since ${\pa_m \psi_m(x) \ov \psi_m(x)}$ dose not decay to zero at any direction of complex infinity. However, the nice observation is that if we once again massage with the Dirichlet boundary condition $\psi_{m\in \mathcal{M}}(0)=0$, we can replace ${\pa_m \psi_m(x) \ov \psi_m(x)}$ with analytic functions $S_\pm(m)$ which have more or less similar asymptotic properties as $s_\pm(\omega)$

\beq \label{eq:master}
~S_+(m)=\frac{2 \left(m+i v \tan \left(\frac{\pi  v}{2}\right)\right) \left(\pi  m^2+v \tan
   \left(\frac{\pi  v}{2}\right) \left(\pi  v \tan \left(\frac{\pi 
   v}{2}\right)+2\right)\right)}{\left(m+i v \tan \left(\frac{\pi  v}{2}\right)\right)^2
   \left(v \tan \left(\frac{\pi  v}{2}\right)+i m\right)-i e^{2  \pi  i m} \left(m-i v \tan
   \left(\frac{\pi  v}{2}\right)\right)^3}
\\S_-(m)=\frac{2 \left(m-i v \tan \left(\frac{\pi  v}{2}\right)\right) \left(\pi 
   m^2+v \tan \left(\frac{\pi  v}{2}\right) \left(\pi  v \tan \left(\frac{\pi 
   v}{2}\right)+2\right)\right)}{i e^{-2 \pi i  m} \left(m+i v \tan \left(\frac{\pi 
   v}{2}\right)\right)^3+ \left(m-i v \tan \left(\frac{\pi 
   v}{2}\right)\right)^2 \left(v \tan \left(\frac{\pi  v}{2}\right)-i m\right)}
\eeq
Given all these ingredients, now we are ready to resum the leading large $N$ four point function for general kinemtical configurations. The steps are the following. First let's enlarge the domain of the parameter $m$ in the sum \eqref{eq:sumf} so that it can be negative (i.e. $\pm m\in \mcal M$, followed by an overall half to remove the double counting). Let's call this summand $f(m)$. Next, usinig the Dirichlet boundary condition once again, we split the summand $f(m)$ as $f(m)\rightarrow f_+(m)+f_-(m)$ such that $f_+(m)S_+(m)$ and $f_-(m)S_-(m)$ have nice asymptotic property at the upper and lower half plane repsectively. It turns out that the expressions $f_+(m)S_+(m)$ and $f_-(m)S_-(m)$ have three additional poles besides $m\in \mcal M$ at $m=\pm v$ and $m=0$.\footnote{We remark that apparent candidate poles at $m=\pm iv \tan({\pi v\ov 2})$ for $S_\pm (m)$ in \eqref{eq:master} disappears when we use the Dirichlet boundary condition in splitting $f(m)\rightarrow f_+(m)+f_-(m)$ in a way that satisfies the required asymptotic properties.} The finally, the Cauchy's theorem gives:

\beq
\mathcal F(x,y;x',y')&=\sum_{\pm m\in \mcal M}f(m)
\\&=\sum_{\pm m\in \mcal M} \left[{1\ov 2\pi i}\oint_{m^{\Circlearrowleft}} f_+(m)S_+(m)+{1\ov 2\pi i}\oint_{m^{\Circlearrowleft}}f_-(m) S_-(m) \right]
\\&= -\left\{\sum_{i=\pm}\text{Res}(f_i(m)S_i(m), 0)+\sum_{i=\pm}\left [\text{Res}(f_i(m)S_i(m), v)+\text{Res}(f_i(m)S_i(m), -v)\right] \right\}
\\&\equiv \mathcal F_{0}(x,y;x',y')+\mathcal F_{v}(x,y;x',y'),
\eeq
where in the final line we define $\mathcal F_0,\mathcal F_v$ as the part of the four point function coming from the Regge poles at $m=0$ and $m=\pm v$ respectively. It turns out that for the TOC case only the $m=0$ Regge pole contributes, while for the OTOC case we have three contributions from $m=0,\pm v$ as illustrated in the figure \ref{fig:Contour}. Our next task is to determine $f_+(m)$ and $f_-(m)$ explicitly for the TOC and OTOC regime and perform a full resummation. 

In the near conformal limit ($\beta J\rightarrow \infty $ or $v\rightarrow 1$) the index set ${\cal M}$ approaches the integers with $\vert m\vert> 1$ which is completely in accord with the absence of the $SL(2)$ zero mode in the near Schwarzian limit \cite{Maldacena:2016hyu,Maldacena:2016upp}. Because of the uniformity of $m$ in the near conformal limit, the resummation can be done easily using the analog of $s_\pm(\omega)$ in the thermal field theory discussed above \cite{Maldacena:2016hyu}.
 In this limit, the four point function is understood as the exchange of a Schwarzian mode between bilocals \cite{Maldacena:2016upp}, and the Sommerfeld-Watson resummation for the Schwarzian propagator in the Lorentzian regime becomes the one with the usual Bose distributions $s_{\pm}(\omega)$ applicable to the Matsubara case, see \cite{Sarosi:2017ykf}.

\subsection{TOC}\label{sec:toc}

Without the loss of the generality, we consider the general kinematical time ordered (TOC) configuration $0\leq \tau_4<\tau_3<\tau_2<\tau_1<2\pi$. We further assume that $0\leq y-y'\leq \pi$, where the rest of the fundamental region $D$ can be determined through the symmetry constraints \eqref{eq:sym}.

Using the procedure explained in sec. \ref{sec:sommerfeld-watson}, we can write explicitly $f^{\text{TOC}}_\pm(m)$ as follows

{\small
\beq
~ f^{\text{TOC}}_+(m)=&{1\ov 8m^2v^4}e^{im(y-y')}\left( im+ v \tan({\pi v \ov 2})\right) 
\\&\left [ -e^{-{1\ov 2}im(x+x')} \left (m+iv \tan({\pi v\ov 2}) \right)^2\left(m-i v \tan({v(\pi-x) \ov 2})\right) \left (m-i v \tan({v(\pi-x') \ov 2}) \right)  \right.
\\&+e^{{1\ov 2}im(x-x')} \left (m^2+v^2 \tan^2({\pi v\ov 2}) \right)\left (m+i v \tan({v(\pi-x) \ov 2})\right) \left (m-i v \tan({v(\pi-x') \ov 2})  \right)
\\&+e^{-{1\ov 2}im(x-x')} \left (m^2+v^2 \tan^2({\pi v\ov 2}) \right)\left(m-i v \tan({v(\pi-x) \ov 2})\right) \left(m+i v \tan({v(\pi-x') \ov 2}) \right)
\\&\left.-e^{{1\ov 2}im(x+x')} \left (m-iv \tan({\pi v\ov 2}) \right)^2\left(m+i v \tan({v(\pi-x) \ov 2})\right)\left(m+i v \tan({v(\pi-x') \ov 2}) \right)  \right]
\eeq

\beq
f^{\text{TOC}}_-(m)=&{1\ov 8m^2v^4}e^{-im(y-y')}\left(-im+ v \tan({\pi v \ov 2})\right) 
\\&\left [ -e^{-{1\ov 2}im(x+x')} \left (m+iv \tan({\pi v\ov 2}) \right)^2\left(m-i v \tan({v(\pi-x) \ov 2})\right)\left(m-i v \tan({v(\pi-x') \ov 2}) \right)  \right.
\\&+e^{{1\ov 2}im(x-x')} \left (m^2+v^2 \tan^2({\pi v\ov 2}) \right)\left(m+i v \tan({v(\pi-x) \ov 2})\right) \left(m-i v \tan({v(\pi-x') \ov 2})  \right)
\\&+e^{-{1\ov 2}im(x-x')} \left (m^2+v^2 \tan^2({\pi v\ov 2}) \right)\left(m-i v \tan({v(\pi-x) \ov 2})\right)\left(m+i v \tan({v(\pi-x') \ov 2}) \right)
\\&\left.-e^{{1\ov 2}im(x+x')} \left (m-iv \tan({\pi v\ov 2}) \right)^2\left(m+i v \tan({v(\pi-x) \ov 2})\right)\left(m+i v \tan({v(\pi-x') \ov 2})  \right) \right]
\eeq }\normalsize

It turns out that the contributions of the Regge pole at $m=\pm v$ is zero in $\mathcal F^{\text{TOC}}$, i.e. $\mathcal F_v^{\text{TOC}}(x,y;x',y')=0$.
Thus all we get the contribution of the $m=0$ Regge pole giving the following time ordered four point function

\beq \label{eq:toc}
~&\mathcal F^{\text{TOC}}(x,y;x',y')
\\&=\mathcal F_0^{\text{TOC}}(x,y;x',y')
\\&=\frac{1}{4 \left(\pi  v
   \tan \left(\frac{\pi  v}{2}\right)+2\right)}
\left(v x \tan \left(\frac{\pi  v}{2}\right) \tan \left(\frac{1}{2} v (\pi
   -x)\right)+2 \tan \left(\frac{1}{2} v (\pi -x)\right)-2 \tan \left(\frac{\pi 
   v}{2}\right)\right)\\&\left(v x' \tan \left(\frac{\pi  v}{2}\right) \tan
   \left(\frac{1}{2} v (\pi -x')\right)+2 \tan \left(\frac{1}{2} v (\pi
   -x')\right)-2 \tan \left(\frac{\pi  v}{2}\right)\right).
\eeq

\subsection{OTOC} \label{sec:otoc}

Here we consider the general kinematical out of time order (OTOC) configuration $0\leq \tau_4<\tau_2<\tau_3<\tau_1< 2\pi$ which automatically satisfies $0<y-y'<\pi$.
We have $f^{\text{OTOC}}_\pm(m)$ as follows
{\small
\beq
f^{\text{OTOC}}_+(m)=&{i\ov 8m^2v^4}e^{-im(y-y')}\left(m-i v \tan({\pi v \ov 2})\right) ^2
\\&\left [ e^{{1\ov 2}im(x+x')} \left (m+iv \tan({\pi v\ov 2}) \right)\left (m+i v \tan({v(\pi-x) \ov 2})\right)\left(m+i v \tan({v(\pi-x') \ov 2}) \right)  \right.
\\&+e^{-{1\ov 2}im(x-x'-4(y-y'))}\left (m+iv \tan({\pi v\ov 2}) \right)\left(m-iv \tan({v(\pi-x) \ov 2})\right)\left (m+i v \tan({v(\pi-x') \ov 2})  \right)
\\&-e^{{1\ov 2}im(x+x'+4(y-y'))}\left(m-i v \tan({\pi v \ov 2})\right)\left (m+i v \tan({v(\pi-x) \ov 2})\right)\left(m+i v \tan({v(\pi-x') \ov 2}) \right)
\\&\left.+e^{{1\ov 2}im(x-x'+4(y-y'))} \left(m+i v \tan({\pi v \ov 2})\right)\left(m+i v \tan({v(\pi-x) \ov 2})\right)\left(m-i v \tan({v(\pi-x') \ov 2})\right)   \right]
\eeq

\beq
 f^{\text{OTOC}}_-(m)=&{i\ov 8m^2v^4}e^{-im(y-y')}\left(m+i v \tan({\pi v \ov 2})\right) ^2
\\&\left [ -e^{-{1\ov 2}im(x+x'-4(y-y'))} \left (m-iv \tan({\pi v\ov 2}) \right)\left (m-i v \tan({v(\pi-x) \ov 2})\right)\left(m-i v \tan({v(\pi-x') \ov 2}) \right)  \right.
\\&-e^{{1\ov 2}im(x-x')}\left (m-iv \tan({\pi v\ov 2}) \right)\left(m+i v \tan({v(\pi-x) \ov 2})\right)\left (m-i v \tan({v(\pi-x') \ov 2})  \right)
\\&-e^{-{1\ov 2}im(x-x')}\left(m-i v \tan({\pi v \ov 2})\right)\left (m-i v \tan({v(\pi-x) \ov 2})\right)\left(m+i v \tan({v(\pi-x') \ov 2}) \right)
\\&\left.+e^{-{1\ov 2}im(x+x')} \left(m+i v \tan({\pi v \ov 2})\right)\left(m-i v \tan({v(\pi-x) \ov 2})\right)\left(m-i v \tan({v(\pi-x') \ov 2})\right)   \right]
\eeq}\normalsize
The contribution to $\mathcal F^{\text{OTOC}}$ from each Regge pole at $m=0$ and $m=\pm v$ are given by

\beq
\mathcal F_0^{\text{OTOC}}(x,y;x',y')=&{-\tan\left ({\pi v \ov 2}\right) \ov 4\pi v \tan\left ({\pi v \ov 2}\right)+8} \left \{ 2v (\pi-2(y-y') \tan\left ({v(\pi-x) \ov 2}\right) \tan\left ({v(\pi-x') \ov 2}\right) \right.
\\&+\tan\left ({\pi v \ov 2}\right) \left [ -4+2v (-\pi+x)\tan\left ({v(\pi-x) \ov 2}\right)+2v (-\pi+x')\tan\left ({v(\pi-x') \ov 2}\right)      \right.
\\&\left. \left.+v^2(-xx'+\pi(x+x'-2(y-y')))\tan\left ({v(\pi-x) \ov 2}\right)     \tan\left ({v(\pi-x') \ov 2}\right)     \right] \right\}
\\ \mathcal F_v^{\text{OTOC}}(x,y;x',y')=&-{1\ov 2}\cos\left ({v({\pi\ov 2}-y+y') }\right)   \sec\left ({\pi v \ov 2}\right) \sec\left ({v(\pi-x) \ov 2}\right)     \sec\left ({v(\pi-x') \ov 2}\right)          
\eeq
The final expression of the leading large $N$ four point function $\mcal F$ is obtained to be

\beq \label{eq:otoc}
~&\mathcal F^{\text{OTOC}}(x,y;x',y')
\\&=\mathcal F_0^{\text{OTOC}}(x,y;x',y')+\mathcal F_v^{\text{OTOC}}(x,y;x',y')
\\&={-\tan\left ({\pi v \ov 2}\right) \ov 4\pi v \tan\left ({\pi v \ov 2}\right)+8} \left \{ 2v (\pi-2(y-y') \tan\left ({v(\pi-x) \ov 2}\right) \tan\left ({v(\pi-x') \ov 2}\right) \right.
\\&+\tan\left ({\pi v \ov 2}\right) \left [ -4+2v (-\pi+x)\tan\left ({v(\pi-x) \ov 2}\right)+2v (-\pi+x')\tan\left ({v(\pi-x') \ov 2}\right)      \right.
\\&\left. \left.+v^2(-xx'+\pi(x+x'-2(y-y')))\tan\left ({v(\pi-x) \ov 2}\right)     \tan\left ({v(\pi-x') \ov 2}\right)     \right] \right\}
\\&-{1\ov 2}\cos\left ({v({\pi\ov 2}-y+y') }\right)   \sec\left ({\pi v \ov 2}\right) \sec\left ({v(\pi-x) \ov 2}\right)     \sec\left ({v(\pi-x') \ov 2}\right)  .
 \eeq

\subsection{Comments on the four point function}

First we remark that the expression of the TOC and OTOC four point function \eqref{eq:toc}, \eqref{eq:otoc} we obtained exactly coincide with the ones obtained by Streicher \cite{Streicher:2019wek}. Also, in the near conformal limit ($v\rightarrow 1$) it reproduces the large $\beta J$ TOC and OTOC correlators in \cite{Maldacena:2016hyu}. In the case of TOC \eqref{eq:toc} we see that $\mathcal F^{\text{TOC}}$ doesn't depend on the $y-y'$ and it factorizes as $\mathcal F^{\text{TOC}}\sim f(x)f'(x)$. This make us  suspect that the disorder averaged operator product expansion of $\lr \psi_i(\tau_1)\psi_i(\tau_2) \rr$ is given by the only conserved quantity of the system, the Hamiltonian; this interpretation in the conformal limit was given in \cite{Maldacena:2016hyu}. The fact that this holds true away from the conformal limit is non-trivial.  \cite{Streicher:2019wek} contains a detailed discussion of this observation.  On the other hand, the OTOC depends on the $y,y'$ and it explicitly confirms the non-maximal Lyapunov exponent $\lambda_L={2\pi \ov \beta}v$ when we take the Lorentzian regime with $\Im(y-y')\rightarrow \infty$, as analyzed in \cite{Maldacena:2016hyu}.

We emphasize that the expressions \eqref{eq:toc}, \eqref{eq:otoc} are not valid along the whole fundamental domain $D~:0\leq x,x'<2\pi , 0\leq y,y'<2\pi$. Clearly $\mathcal F^{\text{TOC}}$ is valid when $0\leq \tau_4<\tau_3<\tau_2<\tau_1<2\pi$\footnote{It is valid for any $0\leq y-y'<2\pi$ because $\mathcal F^{\text{TOC}}$ in \eqref{eq:toc} is independent of $y,y'$.} and $\mathcal F^{\text{OTOC}}$ is valid when $0\leq \tau_4<\tau_2<\tau_3<\tau_1<2\pi$ and these regions do not cover the whole fundamental region.

Let's discuss how to obtain a full four point function on the arbitrary domain. Because of the $2\pi$ periodicity along the Euclidean time direction, it is more transparent to deal with another fundamental region $0\leq \tau_{i}<2\pi$. Once we determine $\mathcal F$ in this domain, we can cover the whole kinematical region using the thermal periodicity. Because of the symmetry \eqref{eq:sym} we can focus on the case of $\tau_1\geq \tau_2$ and $\tau_3\geq\tau_4$. Note that we have already determined the four point function in the region $0\leq \tau_4<\tau_3<\tau_2<\tau_1<2\pi$ and $0\leq \tau_4<\tau_2<\tau_3<\tau_1<2\pi$. Let's denote this region as \textbf{A} and \textbf{B}. Then there are four more regions to consider and the corresponding four point functions can be determined in terms of $\mathcal F_{\textbf{A}}^{\text{TOC}}, \mathcal F_{\textbf{B}}^{\text{OTOC}}$ using the KMS condition:

\begin{itemize}
\item \textbf{C}: $0\leq \tau_2<\tau_4<\tau_3<\tau_1<2\pi$ : TOC

$\mathcal F^{\text{TOC}}_{\textbf{C}}(\tau_1,\tau_2,\tau_3,\tau_4)=\mathcal F_{\textbf{A}}^{\text{TOC}}(2\pi+\tau_2,\tau_1,\tau_3,\tau_4)$

\item \textbf{D}:  $0\leq \tau_4<\tau_2<\tau_1<\tau_3<2\pi$ : TOC

$\mathcal F_{\textbf{D}}^{\text{TOC}}(\tau_1,\tau_2,\tau_3,\tau_4)=\mathcal F_{\textbf{A}}^{\text{TOC}}(\tau_1,\tau_2,\tau_4,\tau_3-2\pi)$

\item \textbf{E}:  $0\leq \tau_2<\tau_4<\tau_1<\tau_3<2\pi$ : OTOC

$\mathcal F^{\text{OTOC}}_{\textbf{E}}(\tau_1,\tau_2,\tau_3,\tau_4)=\mathcal F^{\text{OTOC}}_{\textbf{B}}(\tau_3,\tau_4,\tau_1,\tau_2)$

\item \textbf{F}:  $0\leq \tau_2<\tau_1<\tau_4<\tau_3<2\pi$ : TOC

$\mathcal F^{\text{TOC}}_{\textbf{F}}(\tau_1,\tau_2,\tau_3,\tau_4)=\mathcal F^{\text{TOC}}_{\textbf{A}}(\tau_3,\tau_4,\tau_1,\tau_2)$
\end{itemize}

Finally we provide some non-trivial consistency checks of the expression for $\mathcal F$. First, one can easily check that the expressions for $\mathcal F_{\textbf{A}}^{\text{TOC}}, \mathcal F_{\textbf{B}}^{\text{OTOC}}$ obtained in \eqref{eq:toc}, \eqref{eq:otoc} satisfy the homogeneous equation $\mathcal L \mathcal F=0$ away from the delta function sources in \eqref{eq:green}. Second, at the boundary between the TOC and OTOC regions, it satisfies the relation $F_{\textbf{A}}^{\text{TOC}}-F_{\textbf{B}}^{\text{OTOC}} \vert_{y-y'={x+x'\ov 2}}={1\ov 2}$, which is the right discontinuity enforced by the anticommutation relations between the fermions. It can be shown that this boundary condition between the TOC and the OTOC region is satisfied in the entire  kinematic space. Now these two facts are sufficient to prove that our $\mathcal F$ is indeed a solution of the equation \eqref{eq:green} with the right delta function source. For example, when $(\tau_1,\tau_2)$ is near the source at $(\tau_3,\tau_4)$, we have $\mathcal L \mathcal F \simeq \pa_{\tau_1}\pa_{\tau_2}\theta(\tau_1-\tau_3)\theta(\tau_2-\tau_4)=\delta(x-x')\delta(y-y')$ which corroborates our result.

\section*{Acknowledgements}

We thank  Micha Berkooz, Moshe Rozali, Douglas Stanford, and Gustavo Turiaci for discussions that  piqued our interest in large $q$ SYK, and Mikhail Khramtsov and Hare Krishna Hare Krishna   for initial collaboration. CC is
supported in part by the Simons Foundation grant 488657 (Simons Collaboration on the
Non-Perturbative Bootstrap). MM is supported by the Simons Center for Geometry and Physics. GS was supported in part by the Simons Foundation through the It From Qubit Collaboration (Grant No. 38559) and also by the DOE through Contract No. FG02-05ER-41367 and QuantISED grant DE-SC0020360.


\bibliographystyle{utphys}
\bibliography{refs.bib}

\end{document}